\documentstyle[12pt,times,cite,epsf]{article}
\newcommand{\elltbar}{\overline{\widetilde{\ell}\, }}

\newcommand{\mrm}[1]{\mbox{\rm #1}}

\newcommand{\beq}{\begin{equation}}
\newcommand{\eeq}{\end{equation}}
\newcommand{\bea}{\begin{eqnarray}}
\newcommand{\eea}{\end{eqnarray}}
\newcommand{\eq}[1]{eq.~(\ref{#1})}
\newcommand{\rfn}[1]{(\ref{#1})}
\newcommand{\Eq}[1]{Eq.~(\ref{#1})}
\newcommand{\db}{\hspace{-0.2ex}\not\hspace{-0.7ex}D\hspace{0.1ex}}
\newcommand{\sla}[1]{\hspace{-0.1ex}\not\hspace{-0.5ex} #1\hspace{0.1ex}}
\newcommand{\delte}{\Delta_\epsilon}

\newcommand{\sm}{SU(2)_L \otimes U(1)_Y}

\newcommand{\abs}[1]{\left| #1\right|}
\newcommand{\Tr}[1]{\mathop{\mrm{Tr}}\left\{ #1 \right\}}

\renewcommand{\titlepage}{\clearpage%
\setcounter{footnote}{0}%
\thispagestyle{empty}\pagestyle{plain}\pagenumbering{arabic}%
\kern1mm
\vskip15mm\normalsize}
\newcommand{\docnum}[1]{\hbox to \hsize{\hskip123mm\hbox{#1}\hss}}
\renewcommand{\date}[1]{\hbox to \hsize{\hskip123mm\hbox{#1}\hss}}

\renewcommand{\title}[1]{\vskip1em\begin{center}\Large\bf#1\end{center}\vskip2.5em}
\renewcommand{\author}[1]{\vskip0.5em{\bf #1}\vskip0.5em}
\newcommand{\inst}[1]{\vskip0.3em{ #1}\vskip0.5em}
\renewcommand{\abstract}{\begin{center}{\bf Abstract}\end{center}\quotation}

\begin{document}
\vspace{1cm}
\title{Beyond the Standard Model with Effective
Lagrangians\footnote{ Contribution to the Proceedings
of the 28th Symposium on the Theory of Elementary
Particles, Wendisch-Rietz, August 30 - September 3, 1994, DESY 95-027}}
\begin{center}
\author{Mikhail Bilenky$^a$ and
Arcadi Santamaria$^b$}
\inst{{\bf a}~DESY-IfH, Platanenallee 6, 15738 Zeuthen, Germany \\
and \\
Joint Institute for Nuclear Research, Dubna, Moscow Region, Russia}
\inst{{\bf b}~TH Division, CERN, 1211 Gen\`eve 23, Switzerland \\
and   \\
Departament de
F\'{\i}sica Te\`orica,
Universitat de Val\`encia, and IFIC, Val\`encia, Spain.}
\end{center}
\vspace{0.5cm}

\begin{abstract}
We discuss some applications of the effective quantum field theory
to the description of the physics beyond the Standard
Model. We consider two different examples. In the first
one we derive, at the one-loop level, an effective lagrangian for an
extension of the
Standard Model with a charged scalar singlet
by ``integrating out'' the heavy scalar.
In the second example we illustrate the use
of general effective theories at the loop level.
\end{abstract}

\setcounter{footnote}{0}


If the physical problem contains several distinct energy scales
(masses of the particles etc.) and we are interested in
effects at lower energy scale, then the proper language
is an effective quantum field theory (EQFT) language \cite{georgi}.
In this case the heavy degrees of freedom can be  ``integrated out''
and the physics at lower energy scale can be described by an
effective lagrangian (EL) in the form of the dimensional expansion
\beq
 {\cal L}_{eff}= {\cal L}_0 + \frac{1}{\Lambda} {\cal L}_1 +
 \frac{1}{\Lambda^2}{\cal L}_2+ \cdots~ .
\eeq
Here ${\cal L}_0$ contains operators with canonical dimension
$\le 4$ (which can be renormalizable).  ${\cal L}_n~(n\ge 1)$
are linear combinations of non-renormalizable
operators with dimension $n+4$ which are
suppressed by  $\Lambda^n$,
where $\Lambda$ is an energy scale at which ``new
physics'' starts, and
parametrize our ignorance of the dynamics at
high energies.
Two questions are relevant when the EL
is constructed: \\
- {\it What is the symmetry of the problem?} \\
- {\it What is the (light) particle spectrum?} \\
For any given accuracy physics at energy scale, $E$,
can be described by a limited
number of terms as the contribution of operators of higher dimension is
suppressed by the factor $(E/\Lambda)^n$.
Obviously, when the energy
scale approaches the scale $\Lambda$ one needs more and more terms
in order to describe physics accurately enough.
The renormalizability (in the text-book sense)
is replaced by the requirement
that physics at low scales cannot dramatically depend on the physics
at higher scales \cite{Wei92}.

The EQFT language can be used in two conceptually different cases.
First, when the full
theory is known but it contains heavy degrees of freedom which
can be ``integrated-out'' and one can describe low
energy physics in a very transparent and economic way
with an EL. In this case
parameters of
EQFT are determined completely by the matching to the full theory.
An early example of this case is the lagrangian for low
energy light-by-light scattering derived by Euler and
Heisebrerg \cite{Euler} by integrating
out the ``heavy'' electron in QED.
The second
situation is when the full theory is unknown
and  an  EL  is built only from assumptions about
symmetries and the particle content.
The famous early example of such an approach
is the four-fermion Fermi's theory \cite{Fermi} of weak interaction.
In this case the parameters
of the EQFT can obtained only from experiment.

Although the EQFT approach was used in particle physics for a long
time\footnote{One of the most successful application is the so-called
Chiral Perturbation Theory\cite{ChPT}.}
for the description of electroweak
interactions the main efforts have been made in the direction of the
construction of renormalizable theories by enlarging the symmetry group,
particle content etc.
But the success of the
minimal Standard Model (SM) based on $\sm$
has started to change this point of view. In many recent works
the impact of possible  ``new physics''  is analyzed by
adding to the SM lagrangian effective
non-renormalizable operators
built from the standard fields.
In this case it is  natural to assume the standard $\sm$
symmetry for the new
interactions\footnote{Dimension-six $\sm$ are classified and listed
in \cite{eo}.}.
However, a complication arises due to the
fact that the SM symmetry is spontaneously broken
to $U(1)_{em}$. Then there are two possibilities:

- The gauge symmetry is realized linearly. It means that the Higgs
particle is present in the physical spectrum.
This is the simplest {\it decoupling} situation - effects of
non-renormalizable operators disappear when the scale of ``new
physics'' increases
(only experiment can
tell us something about this scale).
The first term in the EL  is the usual minimal SM.

- The gauge symmetry is realized non-linearly.
There is no elementary scalar in the particle spectrum. The scale
of ``new physics'' cannot be much larger than the Fermi scale as it
has to cure the bad behaviour of the model without Higgs.
Therefore, the operators of higher
dimension become also relevant at the energy scale of modern
experiments. Written in unitary gauge,
the lagrangian has the most
general form consistent with Lorentz invariance and unbroken
$U(1)_{em}$ symmetry.
The first term in the EL is a non-renormalizable
non-linear sigma model\cite{ab,lon}.

The  use of EQFT at tree level is  straightforward.
However, during last years, motivated   by
high precision of the data, people started to bound effective
interactions from their contribution in loops.
This has to be done  with certain caution as such
non-renormalizable interactions give, in general, a divergent result.
Nevertheless, using the appropriate framework one can obtain finite
non-ambiguous results.
As we already mentioned, in the EQFT language
all operators
allowed by the symmetries of the problem
are already present in the EL.
Therefore, there always exists a counterterm available to
absorb any divergence that could appear in loop
calculations.
The price that has to be payed is that
it is not possible to analyze effects of one operator independently
of  other operators that mix with it under renormalization.
Under certain assumptions one can reduce the basis of operators
which mix.
The more assumptions one
makes the stronger will be the bounds one
obtains on the couplings of the effective operators.
The less assumptions one
makes the more reliable will be the bounds obtained.
For example, if we want to analyze an operator that contributes to
experimental observables at one-loop level we can use a ``minimal''
set of operators (which, in general, does not form a closed basis)
that contains the operator in question plus all the operators that
mix ``directly'' with it at the one-loop level.

In this talk we will illustrate the construction
and use of the EQFT by considering two examples: in the first one the
EL is derived from a known underlying model; in the second one
the use of general EQFT at the loop level is discussed.
\vskip 1 cm

First, we consider the construction of the EL from a renormalizable model
which is an extension of the SM with a charged scalar singlet \cite{efflagr}.
The full lagrangian for this  model is
\beq
{\cal L}_{full} = {\cal L}_{SM} + {\cal L}_h,
\eeq
where the ${\cal L}_{SM}$ refers to the minimal SM part and the
${\cal L}_h$ describes the additional charged
scalar singlet, $h$
\beq
{\cal L}_h=
 (D_\mu  h )^\dagger D^\mu h
- m^2 \abs{h}^2 - \alpha \abs{h}^4
- \beta \abs{h}^2 \varphi^\dagger\varphi
+\left(  f_{ab}  \overline{\widetilde{\ell}}_a  \ell_b h^+ +
\mrm{h.c.}\right)~,
\eeq
where the covariant derivative has the form
$D_\mu = \partial_\mu +i g' B_\mu$ (the scalar has hypercharge
$Y=-1$); $\phi$ is a Higgs field, $l$ is a
leptonic $SU(2)$ doublet.

This model is one of the simplest extensions of the SM, but in spite
of its simplicity it has interesting features common to any extension
which contains a large mass scale compared with the Fermi scale
(we assume that $m \ge 1TeV$).
In addition to the coupling of the scalar to leptons, $f$, is
an antisymmetric complex matrix in  flavour space
\cite{efflagr} and this
leads to flavour-changing interactions in the leptonic sector\footnote{
Although the generational lepton number is violated,
the assignment of the total lepton number 2 to the scalar assures that the
total lepton number is conserved; as a consequence neutrinos remain
massless at all orders.}.

If the mass of the scalar, $m$, is much higher than the energy scale of
experiments we can integrate out the scalar. The effective action,
$S_{eff}= \int d^4 x {\cal L}_{eff}(x)$, is defined as
\beq
\label{elnl}
e^{i S_{eff}}  =
e^{i S_{SM}} \int {\cal D} h{\cal D} h^+
\exp\left\{i d^4 x {\cal L}_h(x)\right\}~~,
\eeq
where ${\cal D} h$ represents the functional integration over $h$.
The EQFT represented by the non-local expression \rfn{elnl} is fully
equivalent to the original theory as far as Green functions with
``light'' external particles are considered. As we are interested
in the effects of the heavy scalar ($m \ge 1TeV$) on physics around
the Fermi scale, we will keep only terms of order $O(1/m^2)$.

Expanding (functionally) the full
action around the solution of the classical equation of
motion for the scalar field and integrating over $h$, the
one-loop action can be written (in our approximation) as
\beq
\label{ea}
S_{eff}= S_{SM} +S_h[h_0]+i \Tr{\log(O)}
\eeq
with $O=(-D^2-m^2-\beta \varphi^\dagger \varphi)$.
The last term in \eq{ea} takes into account terms which originate
from the one-loop diagrams with only heavy scalar  in loops. We refer
to \cite{efflagr}
for the details of calculations of the fluctuation
operator and give here the final result which can be
split in two parts.
The first one includes all dimension-six operators
\beq
{\cal L}_{det}^{(1)}=
\frac{1}{m^2} \frac{1}{(4\pi)^2} \left(-\frac{\beta^3}{6}
(\varphi^\dagger\varphi)^3+ \frac{\beta^2}{12}
\partial_\mu (\varphi^\dagger\varphi)\partial^\mu
(\varphi^\dagger\varphi)
+\frac{g'^2\beta}{12} (\varphi^\dagger\varphi) B_{\mu \nu} B^{\mu
  \nu}-
\frac{g'^2}{60} \partial^\mu B_{\mu \nu} \partial_\sigma B^{\sigma\nu}~.
\right)\label{lagr1}
\eeq

The second part of the EL, contains
operators of dimension {\it not larger} than four and they have
ultraviolet (UV) divergent coefficients\footnote{We used dimensional
regularization. Divergences appear as simple poles in $\epsilon$
in the function $\Delta_\epsilon=1/\epsilon-\gamma + 2log(4\pi\mu/m)$.}.
As all $\sm$ operators with dimension $\le 4$ are already present
in the SM lagrangian, this part of the EL  is
absorbed by the redefinition of the SM couplings:
\begin{equation}
\label{barmatchm}
\bar{m}^2_\varphi = m^2_\varphi - m^2
\frac{\beta}{(4\pi)^2}(1+\delte)~,
\end{equation}
\begin{equation}
\label{barmatchlambda}
\bar{\lambda} = \lambda - \frac{\beta^2}{2(4\pi)^2} \delte
\end{equation}
and the $B_\mu$ field
\begin{equation}
\bar{B}_\mu =
\left(1+\frac{g'^2 \Delta_\epsilon}{3(4\pi)^2}\right)^{1/2} B_\mu~.
\end{equation}
In order to keep the canonical form for the covariant derivative
we have to renormalize the hypercharge coupling, $g^{\prime}$,
 as follows,
\begin{equation}
\label{barmatchg}
\bar{g^{\prime}} = \left(1+\frac{g'^2 \delte}{3(4\pi)^2}\right)^{1/2}
g^{\prime}~.
\end{equation}
Note that the above relations are relations between the {\it bare}
couplings of the full lagrangian and the effective one.

Using the equation of motion for the scalar singlet
the second term in the effective action, $S_h[h_0]$,
can be formally written in the following {\it non-local} form
\begin{equation}
\label{lnonlocal}
S_h[h_0] \approx
 -\int d^4x \elltbar(x) f \ell(x)
  \frac{1}{(-D^2-m^2-\beta \varphi^\dagger(x) \varphi(x))}
            \overline{\ell}(x) f^\dagger \widetilde{\ell}(x)\
+O\left(\frac{1}{m^4}\right)~.
\end{equation}
To obtain a {\it local} approximation to it,
 one has to make an expansion
in $1/m^2$:
\begin{equation}
\label{appsolution}
\frac{1}{(-D^2-m^2-\beta \varphi^\dagger(x) \varphi(x))}
= -\frac{1}{m^2}
+ \frac{1}{m^4} (D^2+\beta \varphi^\dagger(x) \varphi(x))+\cdots~.
\end{equation}
Neglecting all terms but the first,
the tree level contribution of the scalar to the EL is:
\begin{equation}
\label{lagr0}
{\cal L}^{(0)} =  \frac{1}{m^2}
 (\elltbar f\ell)
 (\overline{\ell}  f^\dagger\widetilde{\ell})=
 \frac{4}{m^2} f_{ab}  f^*_{a'b'}
 (\overline{\nu^c_{aL}} e_{bL})(\overline{e_{b'L}} \nu^c_{a'L})\ ,
\end{equation}
where the summation over repeated flavour indices
($a,b,a',b'$) is assumed.
It corresponds to the tree-level diagram with scalar exchange between
two lepton currents.

However, by using the expansion \rfn{appsolution} one does not obtain
the complete answer even at order $1/m^2$. Doing this approximation we
assumed that $q^2 \ll m^2$ which is not correct when the scalar
contributes in loops where the loop momentum runs up to infinity.
As a result
we missed many operators
which correspond to one-loop diagrams in the full theory
with heavy-light particles in loops.

In order to find them, we have to consider one-loop diagrams
in the full theory with mixed, heavy-light particles in the loops and
to subtract the corresponding
one-loop contribution
in the effective theory using the tree-level lagrangian \eq{lagr0}
in loops.

In practice, deriving the matching conditions we can avoid the
calculation in the effective theory by splitting the scalar propagator
in two parts
\begin{equation}
\label{splitting}
\frac{1}{k^2-m^2} = -\frac{1}{m^2}+\frac{1}{m^2}
\frac{k^2}{(k^2-m^2)}\
\end{equation}
and using only the second part in calculations. Doing
this we increase the power of the UV-divergence but decrease the power
of the infrared (IR) divergence. Thus, all possible small momentum
singularities, which have nothing to do with the high-energy behaviour,
are transmitted to the low energy EL.
Most of the operators obtained by matching  have
UV-divergent coefficients and serve as counterterms to the divergent
loop contributions that appear in the effective theory, some other operators
have finite coefficients.

To find the operators that appear as a consequence of the matching
procedure we first consider one-loop diagrams
with the minimal number of the external particles
keeping track of the external momenta (to order $p^2/m^2$). Then we
write effective operators which correspond to such amplitudes. After
this we reconstruct the gauge invariance by promoting simple derivatives
to covariant derivatives. Sometimes, however, there is an ambiguity to do
this and we need to calculate diagrams with more external (gauge)
particles.
The full matching procedure can be found in \cite{efflagr}.

As an example, we construct operators
which correspond to the
lepton self-energy with the $h$-scalar in the loop.
In the effective theory this contribution is zero because it is
a massless tadpole-like diagram. However, in the full theory we have a
non-trivial result:
\begin{equation}
\label{selfe}
T_{self-energy} =  \frac{(f^\dagger f)_{ab}}{(4\pi)^2}
\left(2\left(\delte + \frac{1}{2}\right)
+ \frac{2}{3} \frac{p^2}{m^2} \right) \bar{u}(p) \sla{p}
\frac{1}{2} (1-\gamma_5) u(p) \ .
\end{equation}
For the first term in \rfn{selfe} we have the following operator
\begin{equation}
\label{selfenergy}
 2(\delte+\frac{1}{2}) i (\overline{\ell} F \db \ell)\ ,
 \end{equation}
with $F_{ab} \equiv  (f^\dagger f)_{ab}/(4\pi)^2$. Evidently this
operator can be absorbed in the standard kinetic term of the lepton
doublet\footnote{As a consequence we have to redefine the standard
  Yukawa couplings and the coupling of the tree-level four-fermion
  operator, $f_{ab}$.}.

The second term in \eq{selfe} is proportional to $\sla{p} p^2$, which
requires an effective operator of the form
$ i (\overline{\ell}_a \sla{\partial} \partial^2 \ell_b)$.
However, the promotion of this term to covariant derivatives is
ambiguous: should we use $\db D^2$, $\db^3$ or $D_\mu \db D^\mu$ ? The
only way to resolve this ambiguity
is to perform a full calculation with one external gauge
boson \cite{efflagr}.

There are many other effective operators obtained by matching
which correspond to diagrams with
external leptons and Higgs particles and different four-fermion
operators corresponding to box diagrams with a heavy line.


Before presenting the final form of the EL for the model
we discuss the renormalization of the EL.
We use the $\overline{MS}$-scheme
for the renormalization of both the full and
effective theories and, in this case, we obtain matching equations for the
renormalized couplings in both theories \cite{Wei80, Hal81}.
For example, we have the standard relations for the gauge coupling in
the $\overline{MS}$-scheme\footnote{We
denote the $\overline{MS}$ renormalized quantities with the same
symbol as the bare quantities, but adding an additional dependence on the
renormalization scale $\mu$.  All effective theory quantities will be
distinguished by a bar; $D=4-2\epsilon$
and $\frac{1}{\hat{\epsilon}}= \frac{1}{\epsilon}-\gamma+\log
(4\pi)$.}
\begin{eqnarray}
g' \mu^\epsilon &=& g'(\mu)+ \frac{1}{2\hat{\epsilon}}\ b_{g'}
g'^3(\mu)+\cdots\ , \nonumber\\
\bar{g}' \mu^\epsilon &=& \bar{g}'(\mu)+
\frac{1}{2\hat{\epsilon}}\ \bar{b}_{g'} \bar{g}'^3(\mu)+\cdots\
,\nonumber
\end{eqnarray}
where $b_{g'}$ and $\bar{b}_{g'}$ are the lowest-order coefficients of
the
$\beta$-functions for the coupling constants in the full and  the
effective
theories, respectively.  Substituting these equations in
the relation between bare couplings in the full and
effective theories,
\eq{barmatchg}, and equating finite terms we obtain
the desired matching condition for the
{\it renormalized} couplings:
\begin{equation}
\label{matchg}
\bar{g}'(\mu) = g'(\mu) - \frac{g'(\mu)^3}{3(4\pi)^2}
\log(\mu/m)+\cdots\
\  .
\end{equation}

Note that this equation can be obtained by just  dropping the
$1/\hat{\epsilon}\ $ contained in $\delte$ in
\eq{barmatchg}. This is not surprising since the divergent term in
\eq{barmatchg} gives just the charged scalar contribution to the
beta function of $g'$ in the full theory.
For the coefficient of the quadratic term in the Higgs potential we have
\begin{equation}
\label{matchm}
\bar{m}^2_\varphi(\mu) = m^2_\varphi(\mu) - m^2(\mu)
\frac{\beta(\mu)}{(4\pi)^2}(1+2\log(\mu/m))\ ,
\end{equation}
and similar equations can be written for other couplings (and fields).

In order to avoid large logarithms, the matching
conditions should be evaluated at some scale
around the charged scalar mass\footnote{Although,
in principle, they are valid for an arbitrary
value of the renormalization scale $\mu$}.
Then, using the SM
renormalization group, run all the couplings down in order
to obtain their values at lower scales.

\Eq{matchm} is very interesting and a similar equation can be found
in most theories with (at least) two different mass scales.
This equation clearly exhibits the so-called {\it naturalness
  problem} of the SM.
$\bar{m}_\varphi(\mu)$ is the mass parameter that appears in the
Higgs potential part of the effective Lagrangian, and it has to be of
the order of the
electroweak scale. However, if $m(\mu)$ is very
large, one should also take $m_\varphi(\mu)$ large in order to have
$\bar{m}_\varphi(\mu)$ small enough. But even if we do so at some
scale $\mu$, it will be very difficult to keep $\bar{m}_\varphi(\mu)$
small at any other scale. This represents a serious fine-tuning
problem, which appears when the standard model is embedded in
another model containing mass  scales much larger than the Fermi
scale.
It is
important to note that by using $\overline{MS}$-scheme the
problem appears only in the matching conditions.

The final EL in terms of physical fields has the form
(flavour indices are suppressed)
\begin{eqnarray}
{\cal L}^{(1)}
&=&-\frac{g^2}{2 m_W^2} \delta_Z
(c_W^2 J^\mu_A -J^\mu_Z)(c_W^2 J_{A\mu} -J_{Z\mu})
+ \frac{g}{ c _W} \delta_Z Z_\mu (c_W^2 J^\mu_A -J^\mu_Z)\nonumber \\
  &&+\frac{2}{3} \frac{g}{m^2 c_W}
  \left(-(1-2 s_W^2) \left(\delte+\frac{4}{3}\right)+
s_W^2\frac{1}{3}\right)
  \left(M_Z^2 Z^\mu+ \frac{g}{\hat{c}_W} J^\mu_Z\right)
   (\overline{\nu_L} F\gamma_\mu
   \nu_L)  \nonumber  \\
&&+\frac{2}{3} \frac{g}{m^2c_W}
 \left(\left(\delte+\frac{4}{3}\right)+s_W^2\frac{1}{3}\right)
\left(M_Z^2 Z^\mu+ \frac{g}{c_W} J^\mu_Z\right)
 (\overline{e_L} F\gamma_\mu e_L)    \nonumber \\
&&
-\frac{2}{3}\left(\delte+\frac{4}{3}\right)
 \frac{g}{m^2} \left((\sqrt{2} M_W^2 W^+_\mu+J_\mu^\dagger)
   (\overline{\nu_L} F\gamma^\mu
   e_L)+\mrm{h.c.}\right)  \nonumber  \\
&&
-\frac{2}{9} \frac{e^2}{m^2} J_A^\mu (\overline{e} F\gamma_\mu e)\
-\frac{2}{3} \frac{e^2}{m^2} \left(\delte+\frac{5}{3}\right) J_A^\mu
(\overline{\nu_L} F\gamma_\mu \nu_L) \\  \nonumber
&& -\frac{1}{6} \frac{e}{m^2} A^{\mu\nu}
\left((\overline{e_L} F M_e \sigma_{\mu\nu} e_R) +
\mrm{h.c.}\right)\\  \nonumber
 &&+\frac{1}{6}\frac{g}{m^2c_W}
 (1+s^2_W) Z^{\mu\nu}
\left((\overline{e_L} F M_e \sigma_{\mu\nu} e_R)+\mrm{h.c.}\right)
\\   \nonumber
&&-\frac{1}{3\sqrt{2}} \frac{g}{m^2}
\left(W^+_{\mu\nu}
(\overline{\nu_L} F M_e \sigma_{\mu\nu} e_R)+\mrm{h.c.}\right)
\\                       \nonumber
&&-\frac{(4\pi)^2}{m^2}  \left(
(\overline{e_L} F \gamma^\mu e_L) (\overline{e_L} F \gamma_\mu e_L)
+(\overline{\nu_L} F \gamma^\mu \nu_L)
(\overline{\nu_L} F \gamma_\mu \nu_L) \right.\nonumber
\\
&&
\hspace*{1.5cm}
+\left.
2(\overline{e_L} F \gamma^\mu e_L) (\overline{\nu_L} F \gamma_\mu
\nu_L)
\right)~.
 \label{extrabox}
 \end{eqnarray}
 Here $M_e$ is the charged lepton mass matrix and
 $A^{\mu\nu} = \partial^\mu A^\nu-\partial^\nu A^\mu$ and
  $Z^{\mu\nu} = \partial^\mu Z^\nu-\partial^\nu Z^\mu$  are the
 field strengths of the photon and the $Z$ boson, respectively;
$J_A^\mu,~J_Z^\mu,~J_\mu^\dagger$ are the standard
electromagnetic, neutral and charge currents.

This lagrangian shows all phenomenological consequences of the model in
a very transparent way. The most interesting ones are different
processes with generational {\it lepton number violations}. For example,
assuming (without loss of generality) a diagonal form for the
matrix $M_e$, from the fifth line of \rfn{extrabox} we have
the amplitude of the decay
\begin{equation}
\label{ampmuegamma}
T(e_b \rightarrow e_a\, \gamma) = -i \frac{e}{3} F_{ab}
\bar{u}(p_a) \sigma_{\mu\nu} q^\nu (m_b R+m_a L) u(p_b)
\epsilon^\mu(q)
\ ;
\end{equation}
$L$ and $R$ are, respectively,
the left-handed and right-handed chirality operators.
The amplitude \eq{ampmuegamma} leads to the process $\mu \rightarrow
e\gamma$ without neutrino masses.
Other terms lead to decays $\mu^- \rightarrow e^- e^- e^+$ and
similar processes.

Another interesting  process is the flavour changing $Z$-decay,
$Z \rightarrow e^+_a\, e^-_b$. To consider this decay we have to take
into account
not only the contribution at tree level (third line in \rfn{extrabox})
but also the contribution given by the tree-level four-fermion
lagrangian at one loop.
By construction the sum is UV finite and depends
only on the few parameters of the full model.
For example,
from the upper bounds on the branching ratios for
the decays $Z\rightarrow e\mu,~e\tau,~\mu\tau$
measured at LEP \cite{pdg}
one gets $m \ge 1TeV$
(for coupling $f \approx 1$).

However, when the full theory is unknown, the situation is  more
complicated. Assume we are in a two-operator mixing situation
(like in the above case).
Then for their renormalized couplings at some scale $\mu$ we have
\begin{eqnarray}
\label{ces3}
c_1(\mu) &=&
c_1(\mu_0)\left(1+\gamma_{11} \log\frac{\mu}{\mu_0}\right)
+c_2(\mu_0) \gamma_{12} \log\frac{\mu}{\mu_0}\nonumber\\
c_2(\mu) &=&  c_1(\mu_0) \gamma_{21}   \log\frac{\mu}{\mu_0}
+ c_2(\mu_0) \left(1+\gamma_{22}\log\frac{\mu}{\mu_0}\right)\ .
\end{eqnarray}
which are the solutions of the general renormalization group equation
\begin{equation}
\label{renopmix}
\mu \frac{ d c_i(\mu)}{d\mu} = \gamma_{ij} c_j(\mu)\ ,
\end{equation}
valid only in the case that $\gamma_{ij} \log(\mu/m) \ll 1$.
Let us suppose that at experiment we measure the coupling $c_2(\mu)$
at some energy scale $\mu$. If we want to extract bounds on $c_1(\mu_0)$ we
need to know the initial condition, $c_2(\mu_0)$,
as an effective theory predicts only the anomalous dimensions
$\gamma_{ij}$.
Thus, in this case
we need to add $c_2(\mu_0)$  in the
analysis and we either have to consider more experimental data
or make additional assumptions.
As we will see in the next example such an
analysis is more complicated but nevertheless one can get useful
bounds on effective couplings.

\vskip 1cm
One of the most elusive among the non-standard four-fermion
interactions is that which involves only neutrinos. Best bounds on the
effective coupling of the V-A form (we assume lepton universality
for simplicity)
\beq
{\cal L}^{\nu-\nu} = c_1~G_F \sum_{i,j=e,\mu,\tau}
(\bar{\nu}_{i}
\gamma_\alpha (1-\gamma_5)
\nu_{i})
(\bar{\nu}_{j}
\gamma_\alpha (1-\gamma_5)
\nu_{j})\ ,
\label{inter}
\eeq
were obtained \cite{treenunu}
from its tree-level contribution to the invisible width
of the $Z$-boson
via the decay $Z \rightarrow \nu \bar{\nu} \nu \bar{\nu}$:
\beq
|c_1| \le 390 ~.
\label{ourbound1}
\eeq
When the right-handed
  neutrinos are involved in the interaction much stronger bounds were
  obtained recently\cite{masso} from the primordial nucleosynthesis.

Thus, in the case of $V-A$ structure the interaction may be rather strong.
One can ask on the possible bounds one could obtain on this interaction
via its one-loop contribution to the $Z \rightarrow \nu \bar{\nu}$.
Using the fact that the invisible $Z$-width is measured
at LEP with an accuracy better than one-percent  \cite{pdg}
one can get a simple
estimate:
\beq
\frac{\Delta \Gamma_{\bar{\nu} \nu}}{\Gamma_{\bar{\nu} \nu}} \approx
\frac {c_1 G_F M_Z^2}{(4\pi)^2}~~~~~~~~
\rightarrow               ~~~~~~~~~
c_1 \le (1\mrm{--}10)\cdot .
\eeq
This estimate suggests that one can obtain
 good bounds by considering
the four-neutrino operator at the loop level.

The above estimate is rather na\"{\i}ve because
inserting the non-renormalizable
vertex in the loop diagram we get a divergent result.
It should be renormalized by adding a
derivative coupling of the $Z$-boson to neutrinos \cite{nunu}
\beq
{\cal L}^{\nu-Z}=-\frac{g}{2c_W} \mu^\epsilon G_F
\sum_{i=e,\mu,\tau,..}
\left(c_2+\Delta c_2\right)
(\bar{\nu}_{i} \gamma^\alpha L \nu_{i} )
\partial^\beta Z_{\beta\alpha} ~,
\label{ct}
\eeq
where $c_2$ is  the $\overline{\mrm{MS}}$
renormalized coupling and the corresponding counterterm is
\[
\Delta c_2 = - c_1 \gamma_{12} \frac{1}{\hat{\epsilon}}~.
\]
Moreover, since by using only four-neutrino interactions we do
not assume $SU(2)$ symmetry, we have to also add a
non-standard direct (non-derivative)
coupling of the $Z$-boson to neutrinos, $c_3$
(there is no symmetry which forbids it).
Then the full renormalized vertex $Z\bar{\nu}\nu$ is given by
\beq
\label{renamp}
\hat{T} =-\frac{g}{2c_W}
G_F [q^2 \left(c_2(\mu)+
c_1(\mu) \left(\gamma_{12}
\left(\log(\mu^2/|q^2|)+i\pi\theta(q^2)\right)
+ \kappa_{12}\right)\right)+c_3(\mu)]~,
\label{deltag}
\eeq
with
\beq
\gamma_{12}=\frac{1}{3\pi^2}~,~~~\kappa_{12}=\gamma_{12}\frac{17}{12}~.
\eeq
The running couplings in our approximation (we neglect
all contributions with gauge bosons running in the loops) are given by
\bea
&& c_1(\mu) \approx c_1(\mu_0) ~, \\
&& c_2(\mu)=c_2(\mu_0) + c_1(\mu_0) \gamma_{12}
\log \left( \frac{\mu_0^2}{\mu^2}\right)~,
\eea
where $\mu_0$ is some reference scale.
The effective four-neutrino operator
at the one-loop level contributes to the running of the coupling
of the operator \rfn{ct} and we have to consider mixing between at
least
these two operators\footnote{Obviously, there are
many other four-fermion operators like
$(\bar{l}l)(\bar{\nu}\nu)$, etc., which also mix with the $Z$-neutrino
coupling \rfn{ct}. But as we are neglecting loops with gauge bosons,
they do not mix directly at the one-loop level with
the four-neutrino operator and, as they can be strongly bounded
from other processes, we will disregard them.}.
The coupling $c_3(\mu)$ does not mix with the other couplings because it
corresponds to an operator of different dimension, then
$c_3(\mu) \approx c_3(\mu_0)$.

Obviously, we need several experimental data in oder to put bounds on
these couplings in a model independent way.
As the $q^2$ dependence of the coefficients in front of
the various couplings is different, we can separate different
couplings by considering their contribution to the observables at
distinct energy scales: apart from the invisible $Z$-width
measured at LEP we
consider data from high energy deep inelastic scattering (DIS).

The details of our analysis can be found in \cite{nunu} here we list
only final results.
The three-parameter (model independent) fit to the full body of data
gives the following bounds at $68\%$~C.L.
\beq
c_3(\mu_0) =  0.004 \pm 0.009~~~~~~~~
c_2(\mu_0)=  4.7 \pm 7~~~~~~~~
c_1(\mu_0)=  -100 \pm 140~.
\eeq
The extreme values of $c_1(\mu_0)$, of order $\sim 240$, are
possible only because of large
cancellations between the contributions of the
three non-standard couplings.
If one decides that such cancellations are unnatural,
then one obtains a much better bound for the contact four-neutrino
interaction. The complete analysis gives in this case
\begin{equation}
 \abs{c_1(\mu_0)} \le  ~ 2 ~.
\end{equation}
which is  200 times better than bounds on the four-neutrino
coupling from the tree level analysis \eq{ourbound1}.
\vskip 1cm

In this talk we illustrate by two examples
the construction and use of the
effective field theory approach to the description of physics
beyond the minimal Standard Model.
In the first example we sketch the
construction of the one-loop effective lagrangian
for the extension of the Standard Model with a heavy charged
scalar singlet \cite{efflagr}. We discuss the matching of the
effective theory to the underlying full theory.
In the second example \cite{nunu}
we illustrate
the use of the general
effective lagrangian at the loop level by bounding
elusive four-fermion neutrino operator from its contribution
in loops to the invisible width of the $Z$-boson and to the neutral
to charged currents ratio measured in the deep inelastic scattering.


\end{document}